\documentstyle[12pt,epsfig]{article}
\def\beq{\begin{equation}} 
\def\eeq{\end{equation}}
\begin{document} 
\begin{center} 
{\Large{\bf Branched Polymers on the Given-Mandelbrot Family of 
Fractals}}\\[2cm]

{\large{\bf Deepak Dhar}}\\
Department of Theoretical Physics, \\ 
Tata Institute of Fundamental Research, \\ 
Homi Bhabha Road, Mumbai 400~005, INDIA\\ [2cm]

\end{center} 
\bigskip 
\begin{abstract} 

We study the average number $\bar{A}_n$ per site of the number of
different configurations of a branched polymer of $n$ bonds on the
Given-Mandelbrot family of fractals using exact real-space
renormalization.  Different members of the family are characterized by an 
integer parameter $b$, $ 2 \leq b \leq \infty$.  The fractal
dimension varies from $ log_{ _2} 3$  to $2$ as $b$ is varied from $2$ to
$\infty$. We find that for all $ b \geq 3$, $\bar{A}_n$ varies as $
\lambda^n exp( b n ^{\psi})$, where $\lambda$ and $b$ are some constants,
and $ 0 < \psi <1$.  We determine the exponent $\psi$, and the size
exponent $\nu$ (average diameter of polymer varies as $n^\nu$), exactly
for all $b$, $ 3 \leq b \leq \infty$. This generalizes the earlier results
of Knezevic and Vannimenus for $b = 3$ [Phys. Rev {\bf B 35} (1987) 4988].

\end{abstract}

\section{Introduction}

The study of statistical physics models on deterministic fractals has a
long history \cite{nelson, dd1, gefen, rammal}.  Linear and branched
polymers on fractals with finite ramification number provide very simple
and pedagogical examples of renormalization group techniques at work:
these system show a non-trivial critical point, and the values of the
critical exponents can be determined by linearizing the exact real-space
renormalization transformation.  The renormalization equations are coupled
polynomial recursion equations in a finite number of variables, and are
easy to study. By studying different geometrical fractals, one can
investigate how the critical exponents change with the geometrical
properties of the underlying space.

One particular family of fractals which has been used often for such
studies is the Given-Mandelbrot family of fractals \cite{given}. Different
members of the family are characterized by an integer $b$, with $2 \leq b
\leq \infty$. As we increase $b$ from $2$ to $\infty$, the fractal 
dimension increases from
$log_{2} 3$ to $2$.  The critical properties of linear polymers on the
$b=2$ fractal were first studied in \cite{dd1}, and these results were
extended to $ b \leq 8$ by Elezovic et al \cite{elezovic} using the exact
renormalization equations. Surprizingly, it was found that while the
exponent $\nu_b$ appeared to converge to the two-dimensional value $3/4$,
as $b$ was increased from $2$ to $8$, the difference in the susceptibility
exponent $\gamma_b$ from the known exact value $43/32$ in two dimensions
was found to increase with increasing $b$.  This was explained in
\cite{dd3}, where the asymptotic behavior of critical exponents for large
$b$ was determined theoretically using finite-size scaling arguments, and
it was shown that $\gamma_b$ should tend to a different value $133/32$ for
large $b$ .  Numerical Monte Carlo renormalization group techniques have
been used to estimate the critical exponents for significantly larger
values of $b$ up to $ 80$ \cite{mcrg1, mcrg2}.  Knezevic and Vannimenus
(KV) used the real-space renormalization technique to study the properties
of branched polymers on the $b =2$ fractal, and also studied the
transition from the extended phase to collapsed phase \cite{kv1}.  This
was later extended to other fractals, including the $b =3$ fractal
\cite{kv2}.   Dense branched polymers for the $b=2$ fractal have
been studied in the context of spanning trees and loop-erased random walks
\cite{dd4}, and the abelian sandpile model \cite{daerden}.
However, a study of branched polymers on fractals for higher
$b$ has not been undertaken so far. Nor are the properties of the
large-$b$ limit known. 
 
In this paper we study the number of different configurations of an
$n$-bond branched polymer on the Given-Mandelbrot family of fractals using
the exact real-space renormalization group techniques.  On regular
lattices, this number usually varies as $\lambda^n n^{-\theta}$, where
$\lambda$ is some lattice -dependent constant, and $\theta$ is a critical
exponent.  General theoretical arguments that prove the exponential growth
would allow stronger correction terms like $\exp(b n^{\psi})$, with $\psi
< 1$. Why the first correction term to the exponential growth is a simple
power-law term is not fully understood. To see how general is the
power-law correction form, one can study this question on different
graphs, e.g.  fractals.  We find the power-law correction also on the $b =
2 $ fractal. However, this case is exceptional. For all $ b \ne 2$ , while
the number of configurations still increases exponentially with $n$, the
leading correction term to the exponential growth is the
stretched-exponential form:  this number varies as $ \lambda^n e^{b n
^{\psi}}$, where $\lambda$ and $b$ are some constants, and $ 0 < \psi <1$.  
We determine the singularity exponent $\psi$, and the size exponent $\nu$
(average diameter of polymer varies as $n^\nu$), exactly for all $b$, $ 3
\leq b \leq \infty$. This generalizes the earlier results of KV for $b= 2$
and $3$.

This paper is organized as follows: In section 2, we start by
recapitulating the definition of the Given-Mandelbrot family of fractals,
and introduce the generating function for the number of branched polymer
configurations with $n$ monomers. Since the fractal does not have
translational invariance, we average over different positions of the
polymer. The general technique of real-space renormalization applied to
these problems is outlined in Section 3, using the $b = 2 $ case as an
illustrative example. The qualitative behavior of the renormalization
equations for $b \ge 3$ is discussed in Section 4. It turns out that while
the equations involve rather complicated high-degree polynomials, the
critical exponents $\nu$ and $\psi$ do not depend on most of the terms in
these polynomials. We can ignore most of these terms, and still determine
the {\it exact} values of these exponents, if we can identify the
``dominant terms" in the recursion equations. This is done in Section 5.
Finally, in Section 6, using our knowledge of the dominant terms, we
determine the exponents $\nu$ and $\psi$ for all $b \geq 3$, {\it without 
having to write down the full set of recursion equations.}

\section{Definitions}

For any given integer $b$, $2 \leq b < \infty$, the recursive construction
of the Given-Mandelbrot family of fractals is shown in Fig. 1. We start
with a graph with three vertices and three edges forming a triangle. This
is called the first order triangle. To construct the graph of the $(r +
1)$-th order triangle, we take graphs of $b(b + 1)/2$ triangles of $r$-th
order, and glue them together (i.e.  identify corner vertices) as shown in
the figure, to form an equilateral triangle with base which is $b$ times
longer. The case $b=2$ corresponds to the well-known Sierpinski gasket
[Fig.  \ref{sierpinski}].

\begin{figure} 
\begin{center} 
\includegraphics[width=10cm,angle=0]{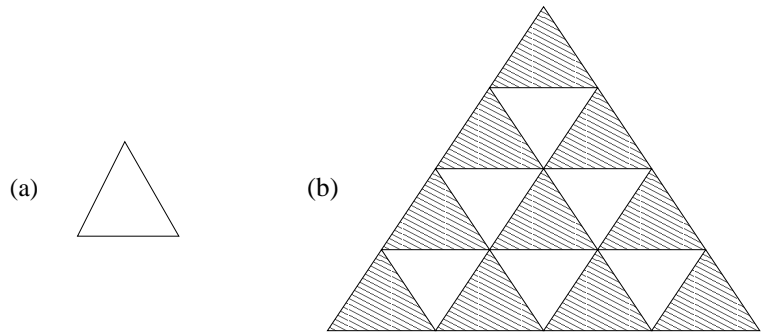} 
\caption{ The recursive construction of the Given-Mandelbrot fractal for 
$b =4$. (a) The graph of  first order riangle.(b) the graph of a $(r +1)$ 
order triangle, formed by joining $b(b+1)/2$  $r$-th order triangles  
shown as   shaded triangles here. } 
\label{given}
\end{center} 
\end{figure}

\begin{figure} 
\begin{center} 
\includegraphics[width=10cm,angle=0]{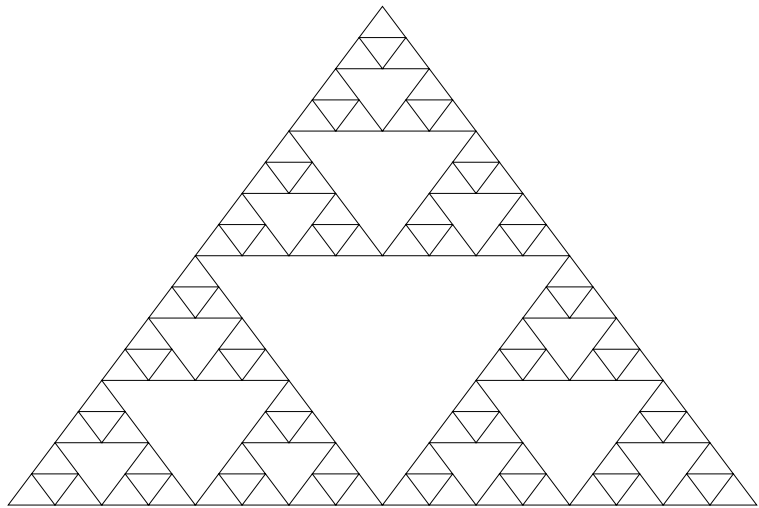} 
\caption{ The  graph of a $5$-th order triangle for $b = 2$. } 
\label{sierpinski}
\end{center} 
\end{figure}

It is easy to see that the number of edges in the graph of the $r$-th
order triangle is \\$3 b^r (b + 1)^r 2^{-r}$, and all vertices have
coordination number $4$ or $6$, except the corner vertices.  The distance
between the corner vertices of $r$-th order triangle is $b^{r-1}$.  Thus,
the fractal dimension of the graph is $ D_b = {\log}_b [b (b + 1)/2]$. For
$b = 2,3, 4..$, these values are $1.5849, 1.6309, 1.6609...$ respectively.  
For large $b$, the fractal dimension tends to $2$ as $D_b \approx 2 -
{\log}_b 2$. The spectral dimension $\tilde{D}_b$ of the graph 
can also be calculated exactly for general $b$  \cite{dd2}. The values of
$\tilde{D}_b$ for $b = 2$ to $10$ are listed in \cite{hilfer}.  For large
$b$, $\tilde{D}_b$ tends to $2$, and the leading correction to its
limiting value is given by $\tilde{D}_b \approx 2 - \frac{\log \log
b}{\log b}$ \cite{hes}.

The determination of the generating function for the branched polymers on
these fractals follows the treatment of \cite{dd1, kv1, kv2}.  Let
$A_n(N)$ be the number of distinct single connected cluster of $n$ bonds
on a graph with $N$ total number of bonds in the graph, different
translations of the cluster being counted as distinct.  For large $N$,
this number increases linearly with $N$. We then define
\beq
\bar{A}_n = Lim_{N \rightarrow \infty} A_n (N) / N;
\eeq
and
\beq
G(x) = \sum_{n=0}^{\infty} \bar{A}_n x^n,
\eeq
where we  assume the convention $\bar{A}_0 = 1$.

\section{Renormalization  equations for the $b=2$ fractal}

\begin{figure}
\begin{center}
\includegraphics[width=14cm,angle=0] {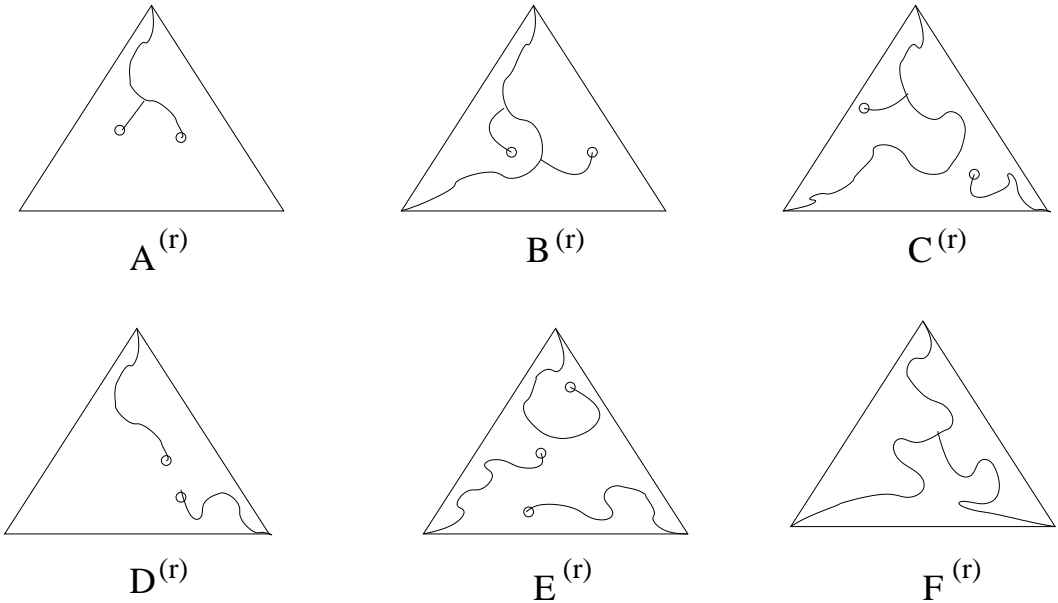}
\caption{ Definition of the restricted partition functions $A^{(r)}, 
B^{(r)}, C^{(r)},D^{(r)}, E^{(r)}$ and $F^{(r)}$. }
\label{abcdef}
\end{center}
\end{figure}

   We assign a weight $x^n$ to each configuration of the polymer with $n$
occupied bonds, and define restricted partition functions $A^{(r)},
B^{(r)}, C^{(r)}, D^{(r)}, E^{(r)}$ and $F^{(r)}$, as shown in Fig
\ref{abcdef}.  Here $A^{(r)}$ is the sum of weights of all connected
configurations of the branched polymer inside a $r$-th order triangle,
such that only one of the corner vertices is occupied by the polymer, and
the other two corner vertices are unoccupied. $D^{(r)}$ is the sum of
weights of all configurations with two mutually disconnected clusters,
each cluster connected to a specified corner vertex. $B^{(r)}$ consists of
sum of all configurations of polymer that connect two specified corner
vertices of the triangle, with the third vertex remaining unoccupied.  
Similarly, $E^{(r)}$ an $F^{(r)}$ correspond to sums over all
configurations with all three corner vertices occupied, and mutually
disconnected, and connected respectively. $C^{(r)}$ corresponds to all
three corner vertices occupied, but only two specified vertices connected
to each other by paths involving occupied bonds lying within the triangle.

The values of these restricted functions for $r =1$ are

\begin{eqnarray}
A^{(1)}=  D^{(1)} =~~ E^{(1)} = 1; ~~B^{(1)}=~~C^{(1)} = x; ~~F^{(1)}= 
3 x^2 + x^3.
\label{eq3}
\end{eqnarray}

 It is straightforward to write down the recursion equations for these 
restricted partition functions at level $(r+1)$ in terms of those at level 
$r$.  One has to sum over all possible polymer configurations for 
different $r$-th order triangles, subject to the constraint that all 
occupied bond are connected to one of the corner vertices of the 
$(r+1)$-th order triangle.  For example, for $b=2$, KV obtained 

\beq
A^{(r+1)} = A[ 1 + 2 B + 2 B^2] + 2  B^2 C + F [ B^2 + A^2 + 2 B D]
\eeq
\beq
B^{(r+1)} = B^2 + B^3 + F [ 4 B C + 2 A B ] + F^2 [ B + D ]
\eeq
\beq
C^{(r+1)} = A B^2 +  3 B^2 C + F [ 7 C^2 + 2 B D] + F^2 [ C + E ]
\eeq
\beq
D^{(r+1)} = A^2 + B[ 6 C^2 + 4 A C + 2 A^2 ] + D [ 2 B + 3 B ^2 ]  +
2 F [ 2 C D + A D + B C + B E]
\eeq
\beq
E^{(r+1)} = A^3 + 14 C^3 + 12 B C D + 6 A B D + 3 B^2 E + 3 F [ C^2 + D^2 
+ 4 C E ]
\eeq
\beq
F^{(r+1)} = 3 F B^2 + 6 F^2 C + F^3
\eeq
where we have dropped the superscript $(r)$ over all the variables in the 
right-hand side of  all the recursion equations to simplify notation.
The generating function $G(x)$ is expressible in terms of these variables
\beq
G(x) = \sum_{r=1}^{\infty} 3^{-r} ( A^{(r)^2} + A^{(r)^2} B^{(r)} + 
B^{(r)^2} D^{(r)})
\label{eq10}
\eeq

It was shown by KV that there exist a critical value $x_c$, such that for
all $x < x_c$, $(A^{(r)}, B^{(r)}, C^{(r)}, D^{(r)}, E^{(r)}, F^{(r)})$
tends to a fixed point $( A^{*}(x), 0, 0, A^{*^2}(x), A^{*^3}(x), 0)$,
where the value of $A^{*}(x)$ increases monotonically from $1$ to $\infty$
as $x$ increases from $0$ to $x_c$. For all $x > x_c$, $A^{(r)}, B^{(r)},
C^{(r)}, D^{(r)}, E^{(r)}, F^{(r)} $ all diverge to infinity. At $x =x_c$,
the values of $A^{(r)}, C^{(r)}, D^{(r)}$ and $E^{(r)} $ diverge to
infinity for large $r$. If we change variables to $\tilde{A} = A F,
\tilde{B} = B, \tilde{C} = C F, \tilde{D} = D F^2 $ and $\tilde{E} = E
F^3, \tilde{F} = F$, the non-trivial fixed point occurs at finite values
of $\tilde{A}, \tilde{B}, \tilde{C},\tilde{D}, \tilde{E}$, with $\tilde{F}
=0$. Linearized analysis of the renormalization equations near the fixed
point determines the singularity exponent $\theta$ for the function $G(x)
\sim |x_c -x|^{\theta -1}$, and the exponent $\nu$ ( correlation length $
\sim |x_c -x |^{-\nu}$). KV found $\nu \simeq 0.71655$, and $\theta \simeq
0.5328$.

\section{Renormalization equations for  $b \geq  3$}

Interestingly, the qualitative behavior of the recursion equations is very
different for $b > 2$.  For $b =3$, the singularity of $G(x)$ is not a
power-law singularity, but an essential singularity \cite{kv2}. The case
$b >3$ has not been studied so far.

For a general value of $b$, the equations are still coupled polynomial
recursion equations in six variables. We denote the six functions
$A^{(r)}, B^{(r)}, \ldots F^{(r)}$ by $K_i^{(r)}$, with $i = 1$ to $6$,
and represent the polynomial recursion equations schematically as
\begin{equation}
K_i^{(r+1)}  = f_i(\{ K_j^{(r)} \}),~~~~\mbox{for} ~~~~i = 1 ~~{\rm to  
}~~ 6. 
\label{eq11}
\end{equation}
where $f_i$ are some ($b$-dependent )  polynomial functions of their six 
arguments. One can also write the generating function $G(x)$ in terms of 
$\{K_i^{(r)}\}$ in a polynomial form similar to Eq.(\ref{eq10}).
\beq
G(x) = \sum_{r=1}^{\infty}~ [b (b+1)/2]^{-r}  g(\{K_j^{(r)} \})
\eeq

It is rather tedious to write down the explicit forms of these polynomials
for any $b > 3$.  The number of terms in each $f_i$ increases very fast
with $b$.  There are approximately 100 terms in each of the recursion
equations for $b = 3$ \cite{knezevic} , and the number would run into
thousands for $b=4$. The number of terms would increase as $ b^{12}$ for
large $b$, as that is the number of polynomials with six variables with
maximum degree $b(b+1)/2$. Also, the coefficients of the terms become very
large, increasing as $exp( b^2)$. Even for $b = 3$, to generate the
recursion equations, one has to use a computer.  It seems clear that a
brute-force approach to determine the recursion equations is not feasible,
except for a few additional values of $b$.

Interestingly, even though the recursion equations are rather complicated,
we show below that for all $ b \geq 3$, the generating function has an
essential singularity of the type $ G(x) \sim exp( a | x_c -x|^{-\alpha} )
$. This corresponds to ${\bar A}_n \sim x_c^{-n} exp( b n^{\psi})$, where
$b$ is some constant, and $\psi = \alpha/(1 + \alpha)$.  We determine the
exact value of the $b$-dependent exponents $\psi$ and $\nu$ for all $b$.

If we start with the initial conditions (Eq. (\ref{eq3})), and iterate the
recursion equations, KV found that many qualitative features of the
behavior of the recursion equations for the $b =3 $ case are same as for
$b=2$ :  for all $x$ below a critical value $ x_c$, $(A^{(r)}, B^{(r)},
C^{(r)}, D^{(r)}, E^{(r)}, F^{(r)})$ tends to a fixed point $( A^{*}(x),
0, 0, A^{*^2}(x), A^{*^3}(x), 0)$, where $A^{*}(x)$ diverges to infinity
as $x$ tends to $x_c$ from below. For all $x > x_c$, the values of all
$\{K_i\}$ tend to infinity for large $r$.  For $x = x_c$, both for $b = 2$
and $3$, the values of $A^{(r)}, D^{(r)}$ and $E^{(r)} $ tend to infinity,
while $ C^{(r)}$ and $ F^{(r)}$ decrease to zero with iteration.  The main
difference is the limiting behavior of $B^{(r)}$ at $x = x_c$.  It tends
to a non-zero limit for $b =2$, but to zero for $b = 3$. Also, the
variables $\{\log K_i^{(r)}\}$ increase linearly with $r$ for $b =2$, but
they increase exponentially with $r$ for $b = 3$.

What makes this problem tractable is the fact that while the polynomial
recursion equations are complicated, the asymptotic behavior of the
variables depends only on a few terms in the recursion equations. To see
this, consider first the simple case when each function $f_i$ has only a
single term, and is of the form 
\begin{equation} 
K_i^{(r+1)} = c_i \prod_{j=1}^{6} [ K_j^{(r)} ]^{{\bf m}_{ij}} 
\label{eq13}
\end{equation} 
where ${\bf m}$ is $6 \times 6$ matrix of non-negative integers. These 
recursion  equations reduce to linear recursions on taking logarithms of 
both sides. We get 
\beq 
\log K_i^{(r+1)} = \sum_{j=1}^{6} {\bf m}_{ij} \log K_j^{(r)} + \log c_i 
\eeq 
These are easily solved. We get that for  large $r$, the leading behavior 
of $K_i^{(r)}$ is given by 
\beq 
\log K_i^{(r+1)} = \sum_{j=1}^{6} [({{\bf m}^r})_{ij} \log K_j^{(1)} + ( 
\frac
{{\bf m}^r -1} {{\bf m} -1})_{ij} \log c_j] 
\eeq 
Let the eigenvalues of  the matrix ${\bf m}$ be $\{ \lambda_{\alpha} \}$, 
$\alpha = 1 $ to $6$,  ordered so that $\lambda_{\alpha} > 
\lambda_{\alpha+1}$. Let the  corresponding right-eigenvectors be $v_{i 
\alpha}$, so that
\beq
{\bf m}_{ij} v_{i \alpha} = \lambda_{\alpha} v_{i \alpha}
\eeq
Then, using the eigenvector decomposition of ${\bf m}$, we see that for 
large $r$ 
\beq
\log K_i^{(r)} = \delta_1 \lambda_1^{r} v_{i1} + \delta_2 \lambda_2^{r}
v_{i2} + {\rm higher~ order~ terms}, 
\label{eq14} 
\eeq 

where $\delta_1$ and $\delta_2$ are some coefficients, that can be
expressed in terms of $K_i^{(1)}$, $c_j$'s, and the left eigenvectors of
${\bf m}$.  As ${\bf m}$ has non-negative elements and $c_i \geq 1, v_{i
1}$ are of the same sign, which can be chosen positive.  Then all $K_i$'s
increase or decrease with iteration for large $r$ according as $\delta_1 >
0$ or $\delta_1 < 0$.  Thus $\delta_1 = 0$ must correspond to $x = x_c$,
and for small deviations of $x$ from $x_c$, by continuity, we will have
$\delta_1$ proportional to $(x - x_c)$.

If $\delta_1 =0$, then the behavior of $K_i$'s for large $r$ is governed
by the second term in Eq.(\ref{eq14}). In this case, all $v_{i 2}$'s are
not of same sign. If $v_{1 2}, v_{4 2}$ and $v_{5 2}$ are positive, and
the rest negative, we would get $K_1, K_4 $ and $K_5$ to diverge, and
$K_2, K_3$ and $K_6$ tend to zero for large $r$, as expected from the 
preceding discussion.

Conversely, consider the full recursion equation(\ref{eq11}).  For any two 
infinite sequences $a^{(r)}$ and $b^{(r)}$ whose logarithm diverges  to 
infinity as  $r$ increases,  we define the notation 
\beq
a \cong b, ~{\rm iff}  ~ \lim_{r \rightarrow \infty} 
\frac{\log a^{(r)} }{\log b^{(r)}} = 1.
\eeq
We make the ansatz  that  at   $x = x_c$,   for large $r$  we have
\beq
K_i^{(r)} \cong exp( v_i \lambda^{r}).
\eeq
Then, as the number of terms in the recursion equations is finite, for 
each  $K_i$, there must be at least one term in the right hand side of its  
recursion equation for which the asymptotic rate of growth of the
logarithm is exactly  as the same as that of the left hand side. We can 
define a matrix ${\bf m}$ such that this dominant term is of the form of 
Eq.(\ref{eq13}). Then we must have
\beq
\lambda v_i = \sum_{j=1}^{6} {\bf m}_{ij} v_j,
\label{eq19}
\eeq 
while all other terms in the equation for $K_i$ are either neglible for 
large $r$, or make a contribution of comparable amount.  These dominant 
terms  define the matrix ${\bf m}$. The vector $\{v_i\}$, then , is a 
right eigenvector of ${\bf m}$, with eigenvalue $\lambda$.

If there are more than one terms that make contributions of the same
order, say in the recursion equation for $K_i$, then for any two such
terms, there are two different row-vectors ${\bf m_i}$ and ${\bf m_i'}$
that satisfy eq.(\ref{eq19}). Then we can subtract these to get a relation
\beq
\sum_{j=1}^{6} [ {\bf m}_{ij} -{\bf m'}_{ij}] v_j = 0.
\eeq

There is one such equation for every such pair. However, all of them are
not independent. We first reduce these to the minimal set of linearly
independent equations. then, each one of such equations can be used to
eliminate one of the $v$'s, and then write Eq.(\ref{eq19}) as an equation
in fewer variables, with a new lower-dimensional matrix ${\bf m}$. This
also changes the coefficients $c_i$ and the eigenvectors, but does not
change the value of $\lambda_1$ and $\lambda_2$. 

For example, for the $b = 3$ fractal, KV found that the dominant terms in 
the recursion equations are 

\beq
A' = 2 A^2 B^2
\eeq 
\beq
B' = A^2 B^2 F + 3 A B^4
\label{eqb'}
\eeq
\beq
C' = 3 A^2 B^4 + A^3 B^2 F
\label{eqc'}
\eeq
\beq
D' = 4 A^3 B^2
\eeq
\beq
E' = 6 A^4 B^2
\eeq
\beq
F' = 2 B^6 + 6 A B^4 F
\label{eqf'}
\eeq

We note that the right hand side involves only $ A, B$ and $F$, and the 
variable $C, D$ and $E$ can be determined in terms of these. 
In Eq.(\ref{eqb'}), the two terms on the right hand side are of same order
iff  $A F \cong  B^2$. This corresponds to the condition
\beq
v_1 -2 v_2  + v_6 =0
\eeq
Note that we get the same condition if we had used Eqs.(\ref{eqc'}), or 
(\ref{eqf'}) instead of Eq.(\ref{eqb'}). Let
\beq
\lim_{r \rightarrow \infty} \frac{A^{(r)} F^{(r)}} {( B^{(r)})^2} = f^{*}
\eeq
  
 Then we can write $F = f^* B^2 /A$ for large $r$, and 
eliminate $F$ and work with a simpler set of recursions 
\beq
A' = 2 A^2 B^2, ~~B' = (3 + f^*) A B^4.
\eeq
This corresponds to a $2 \times 2$ matrix
\beq
{\bf m}= 
\left(\begin{array}{lrlr} 2 &  2 \\ 1 &  4  
\end{array}\right) 
\eeq
With this change of variables, the other equations also reduce to single 
term equations. Also, the matrix ${\bf m}$, and hence the  eigenvalues 
$\lambda_1$ and $\lambda_2$ do not depend on the coefficient $(3 + f^*)$.

Similarly, in other cases, the asymptotic behavior of the variables
$K_i$'s with the full complicated polynomial recursion relations may be
reduced to that for a simpler system of equations where only one term ( to
be called the dominant term) is kept from each polynomial $f_i$ in
Eq.(\ref{eq11}).

 The dominant terms are not unique. Only a few terms in each of the 
polynomials $f_i(\{K_j\})$   are dominant, and any of them can be used for 
determining the matrix ${\bf m}$, and hence  the eigenvalues $\lambda_1$ 
and $\lambda_2$.

We note that in the neighborhood of different fixed points, different
terms are dominant. For example, for branched polymers with attractive
self-interaction, the fixed point corresponding to the dense phase is
$\{K_i\} = \{0, 0, \infty, 0, \infty, \infty\}$.  Clearly, near this fixed
point, the dominant terms are different.

\bigskip

\section{Identifying the dominant terms}

The problem of determining the critical exponents for our problem thus is
reduced to that of identifying what are the dominant terms in the
recursion equations. Each term in the recursion equation corresponds to a
class of configurations of the polymers. However, all possible
combinations of powers are not allowed in a given equation, as there are
connectivity constraints on the allowed configurations. For example, a
term like $E^{\frac {b(b+1)}{2}}$ corresponding to all $r$-th order
triangles with configurations of type $E$ has monomers not connected to
the corners, and is not allowed.

Also, we note that while we may allow polymers with loops, the ratios like
$F^{(r)}/C^{(r)}$ and $B^{(r)}/D^{(r)}$ tend to zero for large $r$. Thus,
any term corresponding to a polymer with loops is dominated by one without
loop, obtained by removing one of the bonds in the loop ( which changes a
type $F$ triangle into type $C$, or type $B$ into type $D$). Thus, loops
are irrelevant, and the dominant terms correspond to polymer
configurations without loops.

To understand  the characteristics of dominant configurations better, it 
is instructive to look at configurations that differ from each other 
locally.

\begin{figure}
\begin{center}
\includegraphics[width=14cm,angle=0] {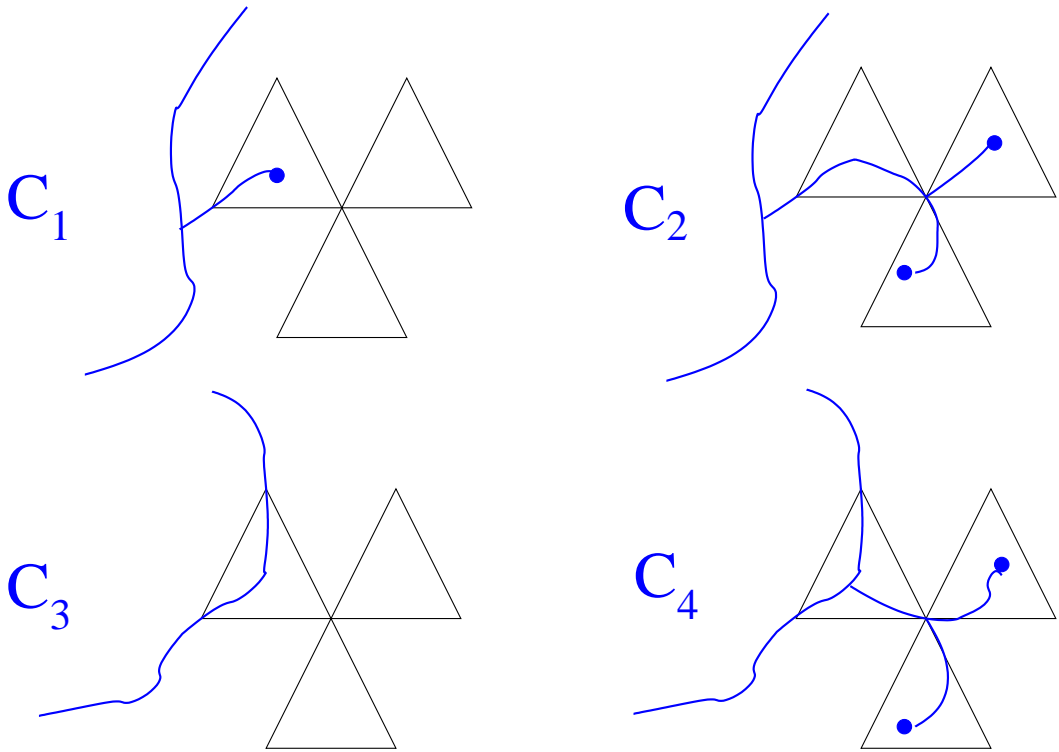}
\caption{ Local modification of polymer confugurations to increase its 
weight. The weight increases is going from configuration $C_1$ to $C_2$, 
and from $C_3$ to $C_4$.  } 
\label{compare1} 
\end{center}
\end{figure}

Consider configurations $C_1$ and $C_2$ shown in Fig. \ref{compare1}. We
assume that these are identical to each other, except in the three $r$-th
order triangles shown. In $C_1$, the vertex common to the three triangles
shown is not connected to the polymer, but in $C_2$ it is. Then if $C_1$
is an allowed configuration, which is connected to the outside in a
specified way, so is $C_2$. If the weight of the rest of polymer outside
the three triangles is ${\mathcal W}$, the weight of $C_1$ is ${\mathcal
W} A$, while that of $C_2$ is ${\mathcal W} B A^2$. The ratio of weights
is $A B$, which tends to infinity for large $r$ ( we shall show this
later). This implies that a configuration of type $C_2$ will dominate over
the corresponding $C_1$.

Again, assume that $C_3$ and $C_4$ in Fig. \ref{compare1} are the same
configuration outside the three triangles shown. The ratio of weights of
$C_4$ and $C_3$ is $ A^2 F / B$.  This also tends to infinity for large
$r$ (also proved later), and hence given a configuration of type $C_1$ or
$C_3$, with two `empty' triangles near the polymer, we can attach a
sidebranch to the polymer to create a configuration that dominates over
it.

\begin{figure}
\begin{center}
\includegraphics[width=14cm,angle=0] {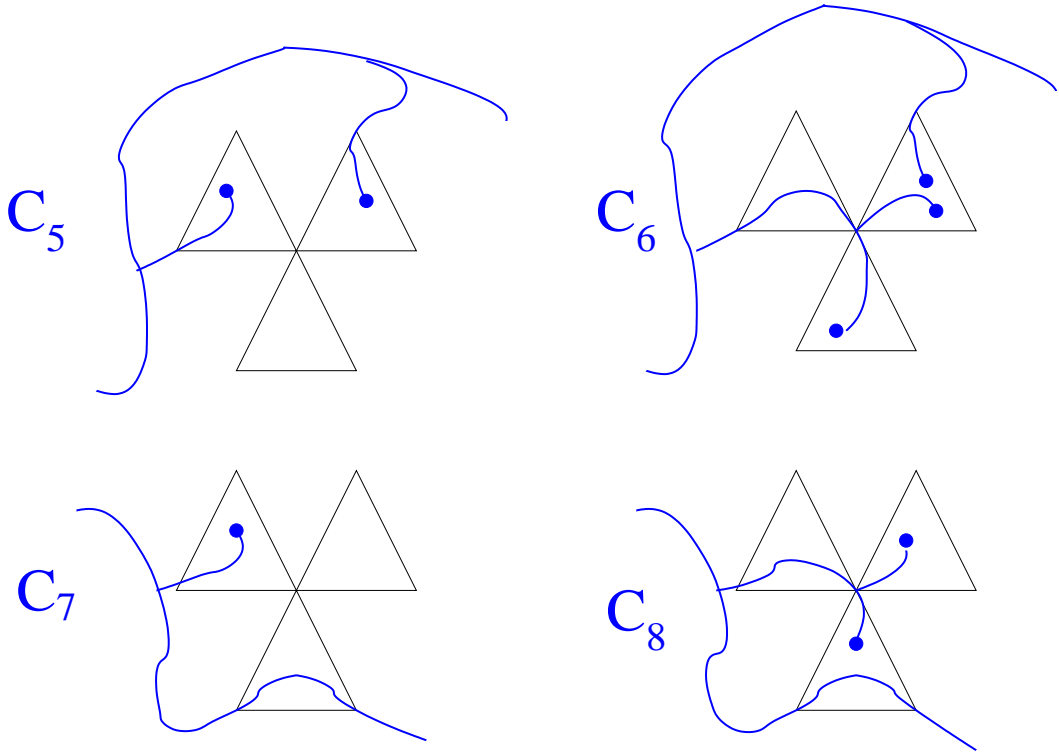}
\caption{Some examples of configurations where extending the polymer 
configuration is not favorable. Here $C_5$ has higher weight than $C_6$, 
and $C_7$ has higher weight than $C_8$.  } 
\label{compare2}
\end{center} 
\end{figure}

This technique of creating a more dominant configuration by attaching
sidebranches works only if the two triangles into which the polymer is
extended are  initially totally empty of polymer. This is shown in Fig.
\ref{compare2}, where again we obtain the configuration $C_6$ by attaching
more monomers to the polymer configuration $C_5$, with configuration of
polymer outside the triangles remaining unchanged. In this case, the ratio
of weights of $C_6$ and $C_5$ is $D B / A$, which we shall show tends to
zero or large $r$. Hence configurations of type $C_5$ dominate over $C_6$.
Similarly, we can see that ratio of weights of $C_8$ and $C_7$ is $C$,
which tends to zero for large $r$. Thus, a configuration like $C_7$ is
dominant over the corresponding configuration of type $C_8$. We see that
$C$ or $D$ type vertices are not favored in creating a dominant
configuration.

\begin{figure}
\begin{center}
\includegraphics[width=11cm,angle=0] {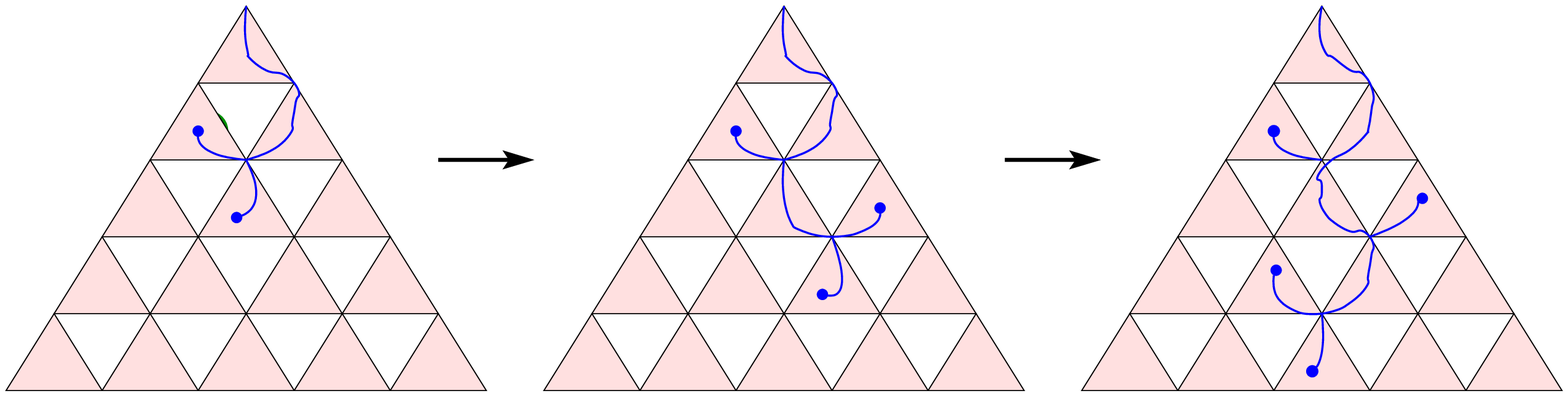}
\caption{ Generating the dominant configuration by extending the cluster.} 
\label{extend}
\end{center} 
\end{figure}

\begin{figure}
\begin{center}
\includegraphics[height=4.5cm,width=12cm,angle=0] {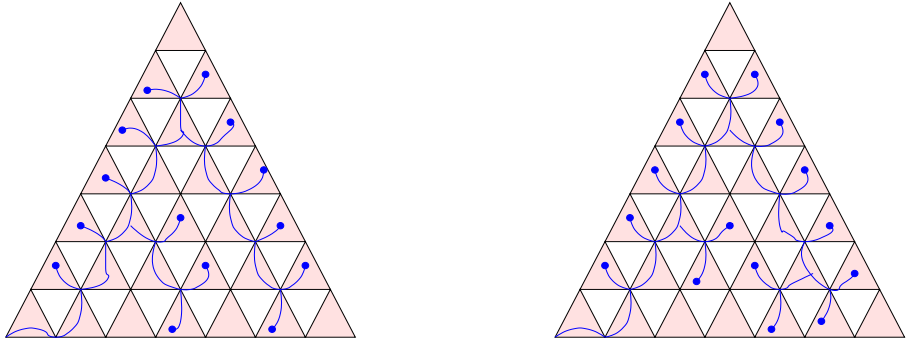}
\caption{Two of the dominant configurations that contribute to $A^{(r+1)}$ 
for the $b = 7$ fractal.  }
\label{b7}
\end{center}
\end{figure}

We can start with any allowed configuration of the polymer, and use this 
local modifications to generate configurations which are more dominant 
(Fig. \ref{extend}). If we start with an initial configuration with a 
small segment of polymer,
with many empty triangles, we find growth at tips or sidebranches is
favorable. We continue this until no such growth sites can be found, and
any further growth of polymer reduces its weight. This is then the maximal
weight configuration. Fig. \ref{b7} shows two such configurations for the
$b =7$ fractal corresponding to configurations of type $A$. We see that in
a maximal weight configuration in a $(r+1)$-th order triangle, the polymer
goes through as many as possible of the corner vertices of the $r$-th
order triangles that are {\it inside} the $(r+1)$-th order triangle ( not
at the boundary of it), and do not contain any type $C$, $D$ or $E$
vertices. This is a generalization of the observation of KV \cite{kv2}
that the dominant terms in the recursion equations had the central node of
the triangle occupied.

\begin{figure}
\begin{center}
\includegraphics[width=14cm,angle=0] {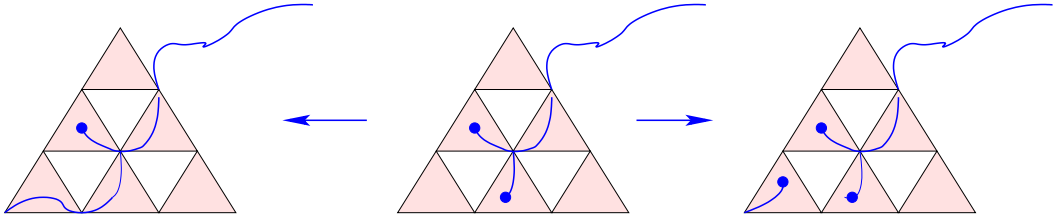}
\caption{Changing a dominant configuration for $A^{(r+1)}$ (middle 
figure) to one for  $B^{(r+1)}$ (left figure)  or $D^{(r+1)}$ (right 
figure) by local change  near a corner.  The polymer is connected to 
another corner vertex of the  $(r+1)$-th order triangle (not shown in 
figure).  } 
\label{atobd}
\end{center} 
\end{figure}

To generate the different terms that are dominant for the different terms
in different $K_i$, we can take the dominant configuration for
$A^{(r+1)}$, and modify it appropriately. For example, consider the
dominant configurations for $B^{(r+1)}$.  Now, in the dominant
configuration for $A^{(r+1)}$, the polymer is as extended as possible, and
hence will reach very close to the other corners of the $(r+1)$-th order
triangle. Then we have to make only a local modification in the triangles
near this corner to connect it to the corner. This is shown in Fig.
\ref{atobd}.  Let the weight of the dominant configurations of $A^{(r+1)}$
be ${\mathcal X}$, and the multiplicative factor needed to convert its
weight into that of $B^{(r+1)}$ be ${\mathcal Y}$.

Then, we have, by definition,
\beq
A^{(r+1)} \cong {\mathcal X}; ~B^{(r+1)} \cong {\mathcal X Y}
\eeq 

To get the dominant configuration of $F^{(r+1)}$, we have to connect the
dominant polymer configuration of $A^{(r+1)}$ to both of the other
corners.   As the polymer in the dominant  configuration reaches close to
both of them, these local modifications can be done independently. Hence
we get 
\beq 
F^{(r+1)} \cong  {\mathcal X Y}^2 
\eeq

To get the dominant configurations for $D^{(r+1)}$, we have to add a 
polymer segment disconnected to the first to the second corner. This can 
also be done by a local modification of the configuration near the corner
( see Fig. \ref{atobd}). We define ${\mathcal Z}$ as the factor by which
we have to multiply the weight of the configuration of $A^{(r+1)}$ to 
affect this change. Thus, we have
\beq
D^{(r+1)} \cong  {\mathcal X Z}
\eeq

Dominant confugurations for $E^{(r+1)}$ involve two such local changes, 
which can be done independently. Similarly, for $C^{(r+1)}$ also, we have 
to make local modifications  at the two corners, connecting one corner to 
the existing $A^{(r+1}$ cluster, and adding a disconnected cluster at the 
second corner. This gives us
\beq
E^{(r+1)} \cong {\mathcal X Z}^2; ~~C^{(r+1)} \cong {\mathcal X Y Z}
\eeq

Thus, as far as the dominant terms are concerned, we can express the six
functions $f_i$ in terms of three functions $\mathcal{X}, \mathcal{Y}$ and 
$\mathcal{Z}$. Equivalently, we can eliminante three of the variables (say
$C, E$ and $F$) using the relations 
\beq
D^2 \cong A E,~~ B^2 \cong A F,~~ A C  \cong B D.
\eeq
  
We note that the weights of the two configurations shown in Fig.  
\ref{b7} are $A^{14} B^{8} F^{2}$ and $A^{15} B^{6} F^{3}$ respectively.
If $ B^2 \cong A F$, these make asymptotically equal contribution $A^{12}
B^{12}$ to ${\mathcal X}$.

Also, as $ A^2 F / B \cong A B$, it follows that the weight increases by
the same factor $A B$ in going from configuration of type $C_1$ to $C_2$,
and from $C_3$ to $C_4$ ( fig. \ref{compare1}). Similarly, the weight
changes by the same factor $B D /A \cong C$, when going from $C_5$ to
$C_6$, as in going from $C_7$ to $C_8$. Again, as $C^{(r)}$ tends to zero
for large $r$, we have justified the asumption made earlier about such
extentions being unfavorable.

\section{Critical exponents for $b \geq 3$}

If we start with the shortest polymer class of configurations that 
contribute to $A^{(r+1)}$ having just weight $A^{(r)}$, then the first 
extension is by a corner vertex at the boundary of the triangle, and the 
resulting weight is $A B$. For any subsequent extention, using either 
extension of type $C_1 \rightarrow C_2$, or $C_3 \rightarrow C_4$, the 
weight increases by a factor $A B$. After $n$ such extentions, the weight 
of the configuration is $A^n B^n$. For weight to be maximal, we want $n$ 
to be as large as possible. It is easy to check that the largest value of 
$n$ allowed (  to be denoted by $\beta$ here) is
\begin{eqnarray}
\beta = (b^2 - 1)/4, {\rm ~if}  ~b {\rm ~is ~odd}, \\
= b^2/4,{\rm ~if } ~b {\rm  ~is ~even.}
\end{eqnarray}

For example, both the configurations shown in Fig.\ref{b7},  have $\beta = 
12$. From the argument given above, any further extension of the polymer 
will be unfavorable. Hence  such configurations correspond to the maximal 
term. If the number of such configurations is  $d(b)$,  we have
\beq
{\mathcal X} = d(b)  A^{\beta} B^{\beta}
\eeq 

It is difficult to determine $d(b)$ explicitly for a general value of $b$.
Of course, one can determine it for small $b$.  This number would be
expected to increase as $exp(b^2)$ for large $b$. Fortunately, its precise
numerical value does not matter for determining the exponents $\nu$ and
$\psi$.

We can similarly determine ${\mathcal Y}$ and ${\mathcal Z}$. To modify 
the dominant configuration of $A^{(r+1)}$ to that for $B^{(r+1)}$, one 
notes that as the former corresponds to a  
maximally extended polymer, one can find specific configurations such that 
the polymer reaches upto the $r$-th order triangle neighboring the corner
which we want to reach. Then only changing the configuration in two such 
triangles is enough (fig. \ref{atobd}).  This gives us
\beq
{\mathcal Y} = B^2 / A
\eeq
Similarly, to get configuration of type $D^{(r+1)}$ from ${\mathcal X}$, 
we need only add an $A$-type  vertex at that corner (fig. \ref{atobd}). 
This gives
\beq
{\mathcal Z} = A.
\eeq
  
Putting all these together, we see that at the critical point, the 
recursion equations for $A^{(r)}$ and $B^{(r)}$ are given by
\beq
\left(\begin{array}{ll} \log A^{(r+1)}\\ \log B^{(r+1)}\end{array}\right) 
= 
\left(\begin{array}{lrlr} ~~\beta &  \beta~~ \\ \beta - 1 &  \beta + 2  
\end{array}\right) 
\left(\begin{array}{ll} \log A^{(r)}\\ \log B^{(r)}\end{array}\right)
+ \left(\begin{array}{ll} c_1\\ c_2 \end{array}\right)
\eeq

For the $2 \times 2$ matrix given above, the equation determining the 
eigenvalues is
\beq
\lambda^2 - 2 ( \beta +1) + 3 \beta =0,
\eeq
whoes roots are
\beq
\lambda_{\pm} = \beta + 1 \pm \sqrt{ \beta ^2 - \beta +1}
\eeq

If $\delta = x_c -x$ is very small, then it increases by a factor 
$\lambda_{+}$  under iteration, till it becomes sufficiently large so that 
linear analysis near the fixed point is no longer applicable. The number of 
iterations $r_{max}$required for the deviation to become ${\cal O}(1)$ is 
given by
\beq
\lambda_{+}^{r_{max}} \delta \simeq {\cal O} (1)
\eeq
whence we get 
\beq
r_{max} = \log(1/\delta) / \log \lambda_{+}
\eeq
the linear size of polymer varies as $b^{r_{max}} \sim (1/\delta)^{\nu}$
with
\beq
1/ \nu =  \log_b \lambda_{+}.
\eeq

It is easy to check that for large $b$, $\nu$ has an expansion of the form
\beq
1/ \nu = D_b (1 -  \frac{1}{ b \log (b^2/2)} + {\rm ~higher~ order~ 
terms}.) 
\eeq

For $\delta \ll 1$, $ log A^{(r)}$  increases as  $\lambda_{-}^{r}$. The 
maximum contribution to $G(x)$ in equation (XX) comes from the term $ r= 
r_{max}$, hence we can replace the sum by the largest term, and write
$log G(x) \sim log {\cal X}^{(r_{max})} \sim \lambda_{-}^{r_{max}}$. This 
gives
$log G(x) \sim (1/\delta)^{-a}$, with
\beq
a = \frac{log \lambda_{-}}{log \lambda_{+}}
\eeq

For large $b$, this varies as $ log(3/2) / log b$, and tends to zero as 
$b$ tends to infinity.

If $G(x)$ varies as $exp[ (x_c -x)^{-a}]$ for $x$ near $x_c$, it is easy
to see that coefficient of $x^n$ in the Taylor expansion of $G(x)$ varies
as $ x_c^{-n} exp( b n^{ \psi})$ where $b$ is some constant, and $\psi =
a/(1 + a)$.

We have listed the numerical  values of the critical exponents $1/\nu$ and  
$\psi$, along with the fractal dimension $D_b$ for some representative 
values of $b$ in Table I.

\section{Discussion}

The critical exponent $\nu$ does not tend to the two-dimensional value as
$b$ tends to infinity. This is not very surprizing, and a similar behavior
has been encountered before in the case of the susceptibility exponent for
linear polymers. Basically, there is a crossover from the two-dimensional
euclidean value to the fractal value.  For polymers with $n$ monomers,
with linear size $\ll b$ ( i.e. $n \ll b^{1/{\bar \nu}}$, where ${\bar
\nu}$ is the 2-dimensional value ), the space looks euclidean, and their
mean size will be similar to that of polymers in regular $2$-dimensional
space. However, polymers with $ n \gg b^{1/{\bar \nu}}$ feel the
constrictions of the corners strongly, and try to avoid them, and become
more compact. Their average size is given by $n^{\nu}$, with critical
exponent $\nu$ dependent on $b$. There is no exponent $\theta$ that we can
define for any $b \ge 3$, because of the presence of the essential
singularity.

One can define the chemical distance exponent $z$ for our problem, just as
we do for the euclidean problem. We take two sites on the branched polymer
at a distance $\ell$ as measured along bonds of the polymer. If the
average euclidean distance between these points is $\bar{r}(\ell)$, we
define the exponent $z$ by the relation $\bar{r}(\ell) \sim \ell^{1/z}$,
for $1 \ll \ell \ll n^{1/\nu}$. We have not been able to calculate $z$ for
different values of $b$.

We note that the logarithm of the number of configurations of polymer acts
like the entropy. The non-translationally invariant fractal lattice
provides a deterministic model for the inhomogenous environment
encountered by the polymer in a random environment. In this case, it would
seem more reasonable to average the logarithm of the number of
configurations of rooted branched polymers over different positions of the
root, as that would correspond to averaging the free energy of the polymer
over different positions of the polymer in space.  Such averages have been
calculated only very recently for linear polymers on fractals
\cite{sumedha}. It would be interesting to see if the stretched
exponentional form for branched polymers is seen also in these ``quenched"  
averages.  Another open problem is the calculation of exponents
characterizing the collapse transition of self-attracting polymers on
these fractals. It is hoped that future works will throw some light on
these questions.

\begin{center}
\vspace{0.5cm}
{\bf Table I} \\
{\bf The fractal dimension $D_b$, and critical exponents $1/\nu$ and 
$\psi$ for some values of $b$}\\
\vspace{0.5cm}

\begin{tabular}{|r|c|c|c|} \hline
$b$  & $D_b$ & $1/\nu$  & $\psi$ \\ \hline
      3  &    ~~ 1.63093~~  &   ~~ 1.41484~~ &    ~~ 0.13250~~  \\
      4  &     1.66096  &    1.55263 &     0.13381 \\
      5  &     1.68261  &    1.57268 &     0.12429 \\
      7  &     1.71241  &    1.64448  &     0.10702 \\
     10  &     1.74036  &    1.70342 &     0.09154 \\
     15  &      1.76787 &    1.74406 &      0.07825 \\
     25  &      1.79685 &    1.78466 &      0.06568 \\
     50  &      1.82788 &    1.82292 &     0.05375 \\
    100  &    1.85165  &     1.84951 &     0.04543 \\ \hline
\end{tabular}

\end{center}

\end{document}